\documentclass{jkas}

\def\year{2023} 
\def\volume{56} 
\def\issue{1} 
\def\beginpage{1} 
\def\received{---} 
\def\accepted{---} 
\def\published{---} 
\date{Received \received; Accepted \accepted; Published \published}

\usepackage{graphicx}
\usepackage{float} 
\usepackage{subfigure}
\usepackage{lineno} 

\def\runningauthor{%
Li et al.
}

\def\runningtitle{%
A Multi-wavelength Study on A Gamma-ray Bright AGN 1308+326 
Using KVN at 22 and 43~GHz
}

\begin{document}
\definecolor{jkasTitleColor}{rgb}{0.02, 0.27, 0.52}
\twocolumn[
\begin{onecolabstract}
{\sffamily {\LARGE \bfseries \noindent \textcolor{jkasTitleColor}{\\\\A Multi-wavelength Study on A Gamma-ray Bright AGN 1308+326\\\\Using KVN at 22 and 43~GHz}}
\\\\
{\large \bfseries{Shan~Li$^{1,2}$, Sang-Sung Lee\orcidlink{0000-0002-6269-594X}$^{1, 2, \star}$, and Whee Yeon Cheong\orcidlink{0009-0002-1871-5824}$^{1, 2}$}}
\\\\
\textcolor{gray}{$^1$ Korea Astronomy and Space Science Institute, Daejeon 34055, Republic of Korea}
\\
\textcolor{gray}{$^2$ University of Science and Technology, Daejeon 34113, Republic of Korea}
\\
$^\star$ Corresponding Author: Sang-Sung Lee, \textcolor{jkasTitleColor}{sslee@kasi.re.kr}
\\\\\\
{\large \bfseries \textcolor{jkasTitleColor}{Abstract}}
\\\\
In this paper, we conduct a multi-frequency analysis of the gamma-ray bright blazar 1308+326 from February 2013 to March 2020, using the Korean VLBI Network at 22 and 43~GHz and gamma-ray data from the Fermi Large Area Telescope (LAT). Our findings reveal spectral variations around the 2014 gamma-ray flare, aligning with the shock-in-jet model. A strong correlation is observed between gamma-ray and 43~GHz emissions, with a 27-day lag in the VLBI core light curve, indicating a 50-day delay from the beginning of a specific radio flare to the gamma-ray peak. This radio flare correlates with a new jet component, suggesting the 2014 gamma-ray flare resulted from its interaction with a stationary component. Our analysis indicates the 2014 gamma-ray flare originated 40-63 parsecs from the central engine, with seed photons for the gamma-ray emission unlikely from the broad-line region.
\\\\
{\bfseries \textcolor{jkasTitleColor}{Keywords:}} galaxies: active --- galaxies: jets --- radio continuum: galaxies --- quasars: general --- quasars: individual:}~(1308+326)
\\\\
\end{onecolabstract}
]

\section{Introduction}
Active galaxies are a special class of galaxies with compact cores experiencing intense physical activities such as high-energy radiation and matter ejection, and these cores are known as Active Galactic Nuclei (AGNs). AGNs emit an excessive amount of non-stellar radiation across all the electromagnetic spectrum. AGNs' tremendous energy output is thought to be powered by supermassive black holes (SMBHs) at their centers~\citep{krolik_active_1999}. SMBHs draw in surrounding gas and dust, forming an accretion disk, generating an extraordinary amount of energy in the form of radiation. Along with this, a fraction of SMBHs also produce relativistic jets—high-speed outflows of charged particles most likely perpendicular to the plane of the accretion disk—that extend over large distances from the central to interstellar space~\citep{blandford_relativistic_2019}. The intense emission created through these processes makes AGNs some of the most luminous objects in the universe. 

Based on their radio loudness, emission lines, viewing angle, and other observed properties, AGNs can be classified into several types. Among these types, blazars are distinguished by their relativistic jets, which are viewed at small angles relative to our line of sight. This characteristic makes them highly variable as a result of relativistic effects. Blazars are further divided into two categories: flat spectrum radio quasars (FSRQs) and BL Lacertae objects (BL Lacs). FSRQs tend to be more luminous and have more powerful jets compared to BL Lacs. Furthermore, FSRQs exhibit broad emission lines in their spectra, while BL Lacs have weak or no emission lines~\citep{urry_unified_1995, padovani_active_2017}.

The spectral energy distribution (SED) of blazars exhibits a distinct pattern with two peaks. The lower-energy peak, spanning from radio to UV-X-rays, is primarily attributed to synchrotron radiation produced by charged particles moving at relativistic speeds in the jet, interacting with magnetic fields. The higher-energy peak, situated in the gamma-ray regime, is most commonly attributable to the inverse Compton (IC) scattering of photons by relativistic electrons within the jet. When the synchrotron photons emitted by the same population of electrons are used as the seed photons for IC scattering, this phenomenon is termed Synchrotron-Self-Compton (SSC). On the other hand, if the primary seed photons originate from sources outside the jet, such as accretion disc radiation, broad-line region (BLR), the dust torus, or the Cosmic Microwave Background (CMB), the process is referred to as External Compton (EC)~\citep[e.g.,][]{prandini2022blazar}. By locating the gamma-ray emission region, we may constrain potential sources of these seed photons, thereby revealing the dominant mechanism of gamma-ray emission.

Research on blazars at various frequency bands with single-dish radio telescopes or very long baseline interferometry (VLBI) has indicated a close relationship between low- and high-energy emission~\citep[e.g.,][]{jorstad_multiepoch_2001,lahteenmaki_testing_2003,taylor_characteristics_2007}. In the Fermi era, advances in the Large Area Telescope (LAT) (improved spatial resolution, larger field of view, and increased sensitivity to lower-energy gamma-rays) significantly enhanced our gamma-ray observation capability, leading to a significant increase in multi-frequency observations. 

\cite{kovalev_relation_2009} carried out a comparative study between the radio emissions of parsec-scale AGN jets and their corresponding gamma-ray properties. This study revealed a strong correlation between the gamma-ray photon flux and the nearly simultaneous compact radio flux density measurements in their radio-selected sample of 135 sources. \cite{max-moerbeck_time_2014} examined the correlation between Fermi/LAT gamma-ray flux and 15~GHz radio flux density of 41 blazars obtained using the 40 m radio telescope of Owens Valley Radio Observatory (OVRO). Their findings indicated that only three sources displayed a correlation with a significance greater than 2.25~$\sigma$. This result led them to suggest that more extended observation periods would be required to achieve higher levels of significance. \cite{meyer_characterizing_2019} employed various methodologies, including a correlation analysis between gamma-ray and radio light curves, to determine the gamma-ray emission region of six Fermi-bright FSRQs. Results from these methods all pointed to a gamma-ray emission region located significantly far from the central engine, well beyond the broad line region.

1308+326~\citep[z=0.996,][]{albareti201713th} is a low-spectral-peaked (LSP) FSRQ with a core-jet structure. It has been observed many times in the past decades in various bands due to its prominent outbursts and high polarization. \cite{puschell_b2_1979} reported results from optical, infrared, millimeter, and centimeter wavelength observations of the source during the spring 1978 outburst, presenting significant changes in its optical-infrared spectral flux distribution, and detected high polarization across optical, infrared, and millimeter wavelengths. However, the relationship between the gradual increase in radio flux density to the optical and infrared activity was not explained. \cite{mufson1983} monitored the same outburst in optical, radio (14.5 and 8.0~GHz), and X- ray bands, highlighting its high variability and activity and suggesting the correlation between the optical and radio outburst is marginal. \cite{tornikoski1994correlated} studied radio and optical variations in a sample of AGNs and found a correlation between different radio frequencies in 1308+326. However, there was no correlation with optical events, emphasizing the need for extensive optical monitoring to understand the connections between rapid optical flares and the flux variations of other frequencies. \cite{watson2000} analyzed the contemporaneous observations at X-ray, optical, and radio wavelengths in June 1996, identifying its quasar characteristics based on high bolometric luminosity, variable line emission, and a high Doppler boosting factor. \cite{hagen2020variability} analyzed the variability of the source using simultaneous optical monitoring results in 2011-2018. They found a strong correlation between optical and gamma-ray emissions, and a polarization direction aligned with the jet direction, indicating a perpendicular magnetic field orientation. In particular, 1308+326 was among 331 AGNs of the Monitoring Of Jets in Active galactic nuclei with the Very Long Baseline Array (VLBA) Experiments (MOJAVE) program~\citep{lister2018mojave}, which have positionally associated gamma-ray counterparts from the Fermi LAT Fourth Source Catalog (4FGL-DR2). A correlation analysis on the source using observational data over decades reviewed that the VLBI radio core of this source lags about 96~days behind the gamma-ray radiation~\citep{kramarenko_decade_2021}.

In this paper, we present the results of simultaneous multi-frequency single-dish and VLBI monthly monitoring observations of 1308+326 from February 2013 to March 2020 (MJD 56350 - 58914) at 22 and 43~GHz. We specifically investigate the potential connection between the variation of the total flux density in the radio range and the gamma-ray outburst in 2014. In Section 2, we describe our observations and data reduction procedures. Section 3 presents the multi-frequency light curves, spectral indices, and cross-correlation results. This is followed by the discussion in Section 4, and finally a summary of the paper is presented in Section 5.

\begin{figure*}[!htb]
\includegraphics[angle=0,width=180mm]{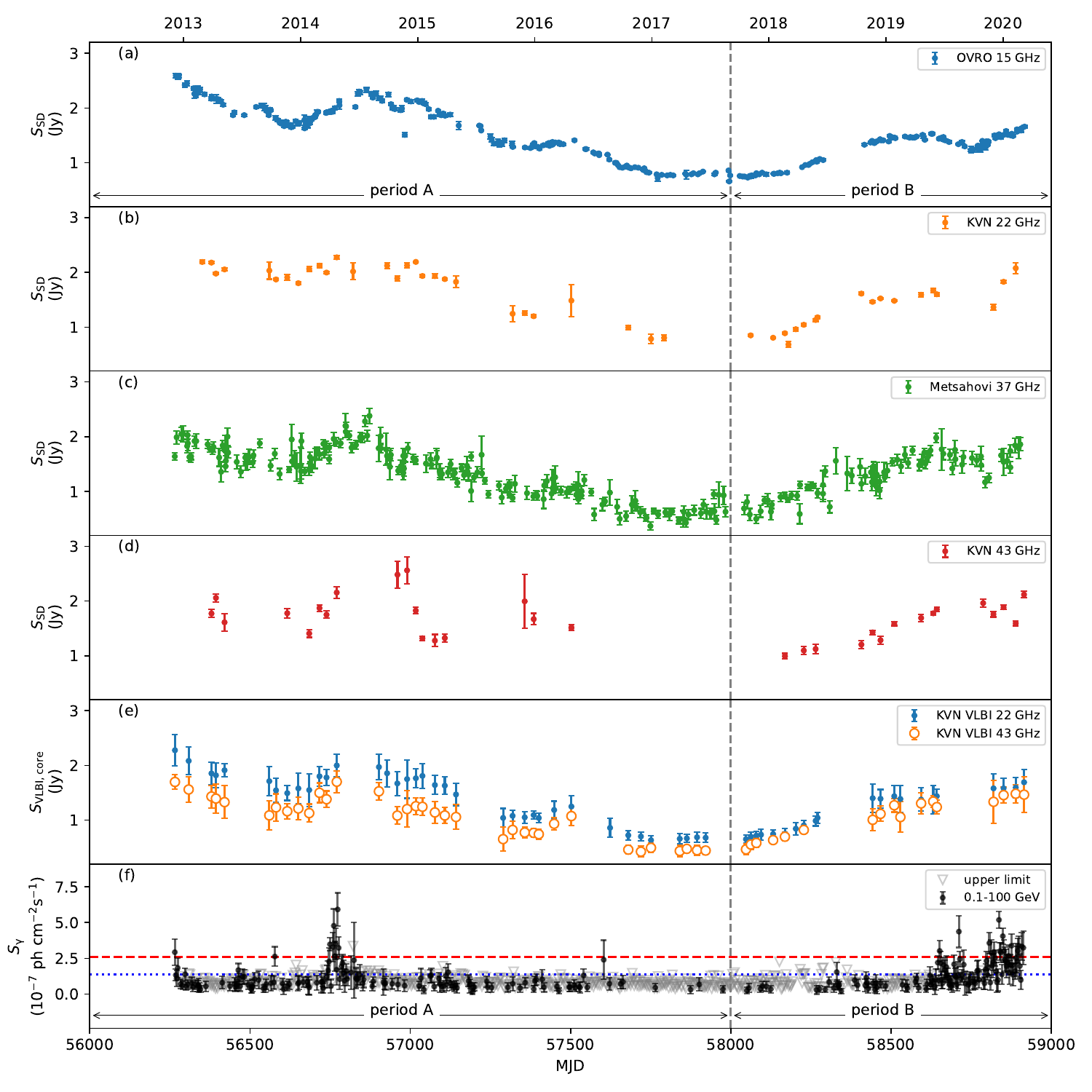}
\caption{
Multi-frequency light curves from 1308+326 observed between Dec 2012 and Mar 2020 (MJD 56265–58914). (a)-(d) are the single-dish flux densities at 15, 22, 37 and 43~GHz. (e) is the KVN VLBI core flux densities at 22, 43~GHz. (f) is the 0.1–100 GeV gamma-
ray light curve. The gray triangles refer to the upper limits of the flux density. The blue horizontal dotted line refers to the active level and the red horizontal dashed line refers to the flaring level (see the text for detail).
The gray vertical dashed line denotes MJD~58000.
 \label{fig:light curves}}
\vspace{0mm} 
\end{figure*}

\section{Observations and Data Acquisition}
\subsection{KVN Data}
\subsubsection{KVN single-dish observations}
1308+326 was included in the list of sources observed by the Korean VLBI Network (KVN) across multi-frequency bands (22–129~GHz) simultaneously, as part of the Interferometric MOnitoring of GAmma-ray Bright AGNs (iMOGABA) program~\citep{lee2016}. This is a Key Science program aimed at studying the origins of the gamma-ray flares in radio-loud AGNs. Observations of iMOGABA sources were conducted monthly with the three 21-m radio telescopes of the KVN, including KVN Yonsei (KYS), KVN Ulsan (KUS), and KVN Tamna (KTN). The full bandwidth of 256~MHz was evenly divided into four frequency bands at 22, 43, 86, and 129~GHz, in single polarization (i.e., left circular polarization)~\citep{lee2011, lee2014early}.

Prior to the VLBI scans (i.e., subsections of the interferometric observations) of individual sources, iMOGABA conducted cross-scans at each antenna to correct pointing offsets. A cross-scan observation produced eight measurements for the four bands in both azimuth (AZ) and elevation (EL) directions. These single-dish cross-scan data were then collected and processed via a Python script pipeline as follows. Each sub-scan yielded a beam pattern represented by the antenna temperature as a function of the relative position angle to the pointing center. A linear function was fitted to the baseline, while a 1-D Gaussian curve was fitted to the main lobe. A Markov chain Monte Carlo (MCMC) method was employed for parameter estimation. The pointing offsets in both the azimuth and elevation directions were corrected using the following equations~\citep{lee2017korean}: 
\begin{flalign}
\label{eq:corrected antenna temperature}
&T_{\rm peak,AZ}^{\rm corrected} = T_{\rm peak,AZ}^{\rm measured} \cdot {\rm exp}\left[ 4 ({\rm ln}{2}) \frac{x_{\rm EL}^2}{\theta_{\rm EL}^2} \right] \\
&T_{\rm peak,EL}^{\rm corrected} = T_{\rm peak,EL}^{\rm measured} \cdot {\rm exp}\left[4 ({\rm ln}{2}) \frac{x_{\rm AZ}^2}{\theta_{\rm AZ}^2}\right] ,&
\end{flalign}
where $ T_{\rm peak,AZ} $ and $ T_{\rm peak,EL} $ are antenna temperatures in Kelvin, $ x_{\rm AZ} $ and $ x_{\rm EL} $ are pointing offsets in arcsecond, and $ \theta_{\rm AZ} $ and $ \theta_{\rm EL} $ are the full width at half maximum (FWHM) in arcsecond of the cross-scan profile fitted with a Gaussian function. We then took the average of the corrected results and obtained the final antenna temperature of one cross-scan.

Next, fitting results with a signal-to-noise ratio (SNR) lower than 5, pointing offsets greater than 20~arcseconds, and FWHM deviated by 30\% from known values (i.e., beam sizes of 122~arcseconds at 22~GHz and 62~arcseconds at 43~GHz) were filtered out to ensure reliability of the results. Finally, the antenna temperature, $T^{*}_{\rm A}$, was converted into flux density, $S_{\nu}$, using the following equation:
\begin{flalign}
\label{eq:flux density}
&S_{\nu} = \frac{2 k T^{*}_{\rm A}}{\eta_{\rm eff}A_{\rm geo}},&
\end{flalign}
where $ k $ represents the Boltzmann constant, $\eta_{\rm eff}$ stands for the aperture efficiency, and $A_{\rm geo}$ represents the geometric area of the telescope. The antenna aperture efficiencies can be found on the KVN homepage (\url{http://kvn.kasi.re.kr}). The inverse variance weighted mean was computed for multiple scans per epoch, and then the flux densities from the three antennas were averaged to obtain the epoch average flux densities. The mean uncertainties of the flux density measured at 22 and 43~GHz are 0.05~Jy and 0.09~Jy, respectively.

\subsubsection{KVN VLBI data}
The VLBI data of the KVN observations on 1308+326 were correlated by the DiFX correlator and reduced by the KVN pipeline~\citep{hodgson2016_kvnpipe}, which has been further modified to include ionospheric delay corrections derived from total electron content maps, updated Earth orientation parameter corrections, and parallactic angle corrections.
The calibrated data have been used to obtain the core flux density of the source at 22 and 43~GHz using a script developed by Cheong et al. (in prep) which uses closure quantities and visibility amplitudes for fitting the source brightness with multiple 2-dimensional Gaussian models. The uncertainty determined from the fitting process were substantially lower than expected ($\sim1\%$), indicating that they did not accurately reflect the true uncertainty of the data. Consequently, we followed the method described in section 2.4 of \cite{lee2016} for calculating flux density uncertainty. This approach required us to obtain the root mean square (RMS) of the residual map, for which a phase-only self calibration was performed after deriving the optimal source brightness model for each epoch. This additional calibration step was essential for deriving a more accurate and representative measure of the flux density uncertainty. The mean uncertainties of the core flux density measured at 22 and 43~GHz are 0.17~Jy and 0.16~Jy, respectively.

\subsection{Multi-frequency Data}
\subsubsection{OVRO 15~GHz and Metsähovi 37~GHz}
1308+326 has been observed at 15~GHz with the OVRO 40~meter radio telescope as a part of the OVRO blazar monitoring program since 2008. Complete details of the reduction and calibration procedure are given in \cite{richards2011blazars}. 

The 37~GHz observation of the source was made with the 13.7~meter Metsähovi radio telescope. This radio telescope has been used to monitor hundreds of AGNs since the early 1980s. A detailed description of the data reduction and analysis is given in \cite{Teraesranta1998AFifteen}.

A part of the flux density data of the source at both wavelengths were published in \cite{hovatta2021association}. For our analysis, we have extracted data spanning from December 2012 to March 2020 (MJD~56268–58916) and observed a mean observation cadence of about 9~days at 15~GHz and about 10~days at 37~GHz. This denser observed data was utilized for comparison with the more sparsely sampled KVN single-dish data, aiding in our analysis of the KVN data.

\subsubsection{Gamma-ray data}
The gamma-ray data of 1308+326 are obtained from the Fermi Large Area Telescope (LAT) light curve repository \footnote{\url{https://fermi.gsfc.nas a.gov/ssc/data/access/lat/LightCurveRepository/}} (LCR, \citealp{abdollahi2023}). The LCR is a database that contains flux-calibrated light curves for over 1500 variable sources listed in the Fermi LAT point source catalog. These light curves span the entire Fermi mission, are regularly updated, and provide data at various time intervals (e.g., 3~days, 1~week, and 1~month). We have selected the 3-day-cadence photon flux of 1308+326, analyzed with an energy bin of 0.1–100~GeV and a fixed spectral index. The minimum detection threshold (the upper limit) is 2$\sigma$. 

\subsubsection{VLBA data}
1308+326 is monitored by the Boston University group at 43~GHz under the VLBA-BU-BLAZAR monitoring program\footnote{\url{https://www.bu.edu/blazars/BEAM-ME.html}}, as part of a sample of 38 gamma-ray bright blazars. Each source has been detected in gamma-ray energies by the Fermi LAT, with an average flux density at 43~GHz exceeding 0.5~Jy. Observations are carried out roughly monthly via dynamic scheduling, and the methodology for data processing is described in \cite{jorstad2017} and \cite{weaver2022kinematics}. Comprehensive results of these observations, spanning from June 2007 to December 2018, have been presented in \cite{weaver2022kinematics}. For the purposes of this study, we have extracted analysis results concerning the parsec-scale jet kinematics of 1308+326 from January 2013 to December 2018. These data have been replotted for comparative analysis with our own observational findings.

\section{Results and analysis}

\subsection{Multi-frequency Light Curves}
\label{sec:lc}
Figure~\ref{fig:light curves} illustrates the multi-frequency light curves of 1308+326, obtained between December 2012 and March 2020 (MJD~56265-58916), using the KVN at 22, 43~GHz, OVRO at 15~GHz, Metsähovi at 37~GHz and the Fermi LAT at the gamma-ray bands.

\subsubsection{Radio light curves}
\label{sec:radio light curve}
Figure~\ref{fig:light curves} (b) and (d) present light curves from KVN single-dish observations at 22 and 43~GHz. These were observed simultaneously with a mean cadence of 56~days at 22~GHz and 79~days at~43~GHz. Occasional large time gaps are present due to system maintenance, poor weather, and other factors. The data sets include 46 flux measurements at 22~GHz and 33 at 43~GHz. The total flux densities range from 0.68 to 2.27~Jy at 22~GHz, and from 0.99 to 2.56~Jy at 43~GHz during the entire observation period.

The light curves can be divided into two distinct periods based on the observed flux density: period A, before MJD~58000 (4th September 2017), and period B, after this date. During period A, there appears to be a general decline in fluxes, with noticeable peaks observed in April 2014 at 22~GHz and in November 2014 at 43~GHz. During period B, the fluxes begin to resurge with some fluctuations, culminating in a significant rise towards the end of the data points in both radio bands.

A comparable trend is also discernible in the VLBI core light curves, as illustrated in Figure~\ref{fig:light curves} (e). However, this trend appears smoother due to the increased density of data points. The VLBI core flux densities range from 0.63 to 2.28~Jy at 22~GHz, and from 0.43 to 1.70~Jy at 43~GHz during the entire observation period. 

We found that while single-dish flux densities at 43~GHz noticeably exceed those at 22~GHz during certain periods (i.e., MJD 56960, 56990, 57357, 57385), this trend is absent in the VLBI core light curves where flux densities in both bands fluctuate in unison. Comparing these light curves with those from OVRO (Figure~\ref{fig:light curves} (a)) and Metsähovi (Figure~\ref{fig:light curves} (c)), we noticed that the observed values for these specific epochs in the other two frequencies did not exhibit a flux density at higher frequency surpassing that at the lower one. When we backtracked the original measurements for the 43~GHz single-dish observation, we found that for each of these four epochs, there was only a single measurement from one antenna in each of the azimuth and elevation direction, preventing the reduction of random errors by averaging. In contrast, data from other epochs had at least two measurements, and the corresponding data at 22~GHz contained a minimum of three measurements, resulting in significantly reduced errors. Additionally, the antenna temperature of these epochs exhibited extremely high residuals after Gaussian fitting. When converted to flux density, the RMS of the residuals ranged from 0.31 to 0.68~Jy. Although we have excluded results with an SNR less than 5 and pointing offsets larger than 20~arcseconds post-fitting, such systematic gain fluctuations may further amplified the already significant errors, undermining the reliability of these data. This factor should be taken into account in subsequent analyses. The RMS of the model fitting residuals from measurements taken on these four epochs, as well as from epochs close to them, is presented in Appendix \ref{Append_A}.

\subsubsection{Gamma-ray light curve}
The gamma-ray light curve, presented in Figure~\ref{fig:light curves} (f), was sampled at a mean cadence of 10.37~days, excluding the upper limits (triangle markers). The flux density of the source at gamma-ray band varies considerably, from $1.53\times 10^{-8} \rm{~ph} \rm{~cm}^{-2} \rm{~s}^{-1}$ to $5.89\times 10^{-7} \rm{~ph} \rm{~cm}^{-2} \rm{~s}^{-1}$, over the observed time-frame. Prominent peaks are notably observed in early 2014 and late 2019, suggesting sustained activities during these corresponding periods. Intervals of relatively low gamma-ray flux (e.g., 2017-2018) indicate periods of reduced activity or quiescence.

Active states and flares in the gamma-ray light curve are defined based on the weighted mean flux, $\langle F_w \rangle$, and its weighted standard deviation, $\sigma_{w}$. A lower limit to an active flux level is defined as $\langle F_w \rangle + 1\sigma_w$, and the threshold of a flaring flux level is defined as $\langle F_w \rangle + 3\sigma_w$~\citep{williamson2014comprehensive}.

We observed an apparent correlation between the gamma-ray flares in early 2014 and the radio flaring periods. In the single-dish 43~GHz light curve, there was a sharp rise (MJD~56989) and fall in flux approximately 200~days after the gamma-ray outburst, which occurred from April 2nd to 26th, 2014 (MJD~56749-56773). The flux density of both 22 and 43~GHz VLBI core also rose during MJD~56770--56902 after the gamma-ray flare.

\begin{figure*}[!htb]
\includegraphics[angle=0,width=170mm]{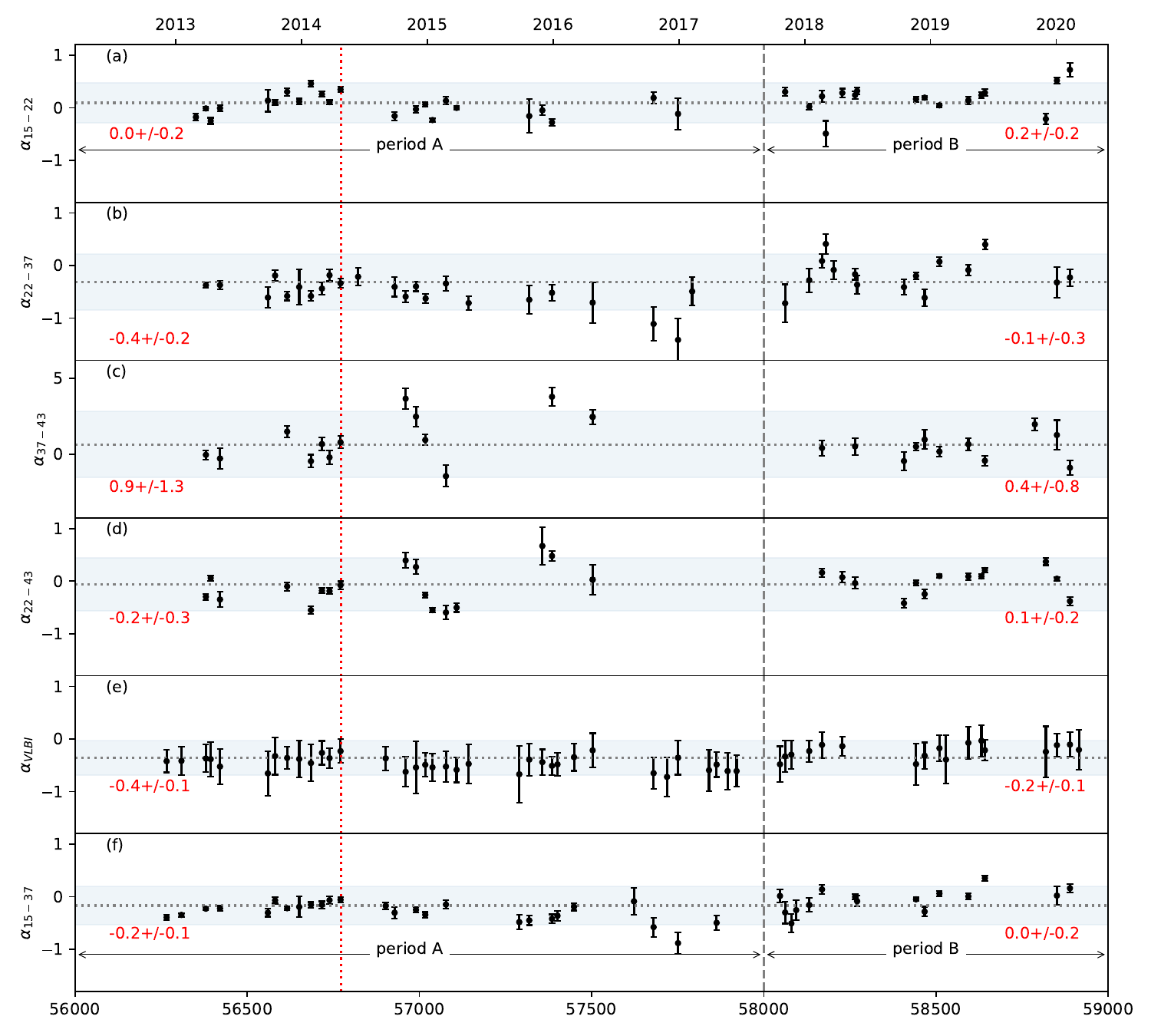}
\caption{Radio spectral indices measured between single-dish radio flux density at 15-22~GHz (a), 22-37~GHz (b), 37-43~GHz (c), 22-43~GHz (d), and 15-37~GHz. (e) spectral indices measured between VLBI core flux density at 22-43~GHz. The gray horizontal dotted line refers to the weighted average over the entire period. The light-blue areas indicate 2$\sigma$ significance level. The red vertical line denotes the peak time of the gamma-ray flare. The gray vertical dashed line denotes MJD~58000. The red numbers denote the weighted averages and their 1$\sigma$ uncertainty in each period.
 \label{fig:spectral index}}
\vspace{0mm} 
\end{figure*}

\subsection{Spectral Indices}
To investigate the spectral property and its variation, we measured the spectral indices for 22 and 43~GHz single-dish as $\alpha_{22-43}$, and VLBI core light curves as $\alpha_{\rm VLBI}$, using a power law, $S_{\nu} = \nu^{\alpha}$ (Figure~\ref{fig:spectral index} (d), (e)),
\begin{flalign}
\label{eq:spectral indices}
&\alpha_{22-43} = \frac{{\rm log} S_{43} - {\rm log} S_{22}}{{\rm log} \nu_{43} - {\rm log} \nu_{22}},&
\end{flalign}
where $S_{22}$ and $S_{43}$ are the single-dish or VLBI core flux densities at 22 and 43~GHz, respectively.

The spectral indices derived from 22 and 43~GHz single-dish flux densities exhibited variability throughout the observation period, with values ranging from $-0.59$ to $0.67$. The weighted average and standard deviation were calculated as $-0.2 \pm 0.3$ during period A, and $-0.1 \pm 0.2$ during period B. This indicates that, on average, the source exhibited a flat spectrum over the observed period, with significant variations that occasionally displayed either inverted or steep spectral characteristics.  Several spectral indices were significantly influenced by the relatively large errors associated with the 43~GHz single-dish flux density measurements (i.e., MJD~56960, 56990, 57357, 57385), reaching values around $0.5$. However, even excluding these data points, the upward trend preceding the gamma-ray flare and the decline thereafter remain noticeable. Such variations suggest temporal changes in the physical conditions within the jet, for instance, fluctuations in particle density.

On the contrast, the fluctuations in 22 and 43~GHz VLBI core spectral indices were less pronounced. Throughout the entire observation period, the values consistently remained negative, ranging from $-0.72$ to $-0.03$, with the maximum value during period A being about $-0.37$. This indicates that the VLBI core predominantly exhibited flatter spectra during most of the observation period.

Given the relatively large average error margin for $\alpha_{\rm VLBI}$ ($\sim76\%$), which is approximately twice that of $\alpha_{22-43}$ ($\sim39\%$), we also calculated spectral indices $\alpha_{15-22}$, $\alpha_{22-37}$, $\alpha_{37-43}$, and $\alpha_{15-37}$ for comparison, using single-dish flux densities from OVRO and Metsähovi. OVRO and Metsähovi provided weekly observational data, which is denser than monthly observations from KVN. Therefore, we treated the average flux from OVRO and Metsähovi within a 7-day window around the KVN observation dates as quasi-simultaneous for calculations. The resulting spectral indices, depicted in Figure~\ref{fig:spectral index} (a), (b), (c) and (f), indicate that, excluding the influence of the single-dish 43~GHz flux density, an upward trend prior to the gamma-ray flare (optically thick) and a downturn thereafter (optically thin) also exist in the other bands.

To further quantify the consistency and reliability of our observations, we calculated the correlation coefficients between different spectral indices. The results showed that the VLBI spectral indices had a correlation coefficient of less than or equal to 0.2 with all single-dish spectral indices, while some single-dish spectral indices showed correlation coefficients between 0.5 and 0.7 ($\alpha_{15-37}$ and $\alpha_{15-22}$, $\alpha_{15-37}$ and $\alpha_{22-37}$, $\alpha_{22-43}$ and $\alpha_{37-43}$). Furthermore, the presence or absence of data points from problematic epochs in 43~GHz single-dish light curve had little effect on the calculation of the correlation coefficients. This suggests that the variability observed in the single-dish spectral indices may be intrinsic, and the spectral evolution (optically thin/thick transition) before and after the 2014 gamma-ray flare is evident across the 15-43~GHz range. The less pronounced variation in VLBI spectral indices is largely due to their higher uncertainty ratio.

\begin{figure*}[!htb]
\centering
\includegraphics[angle=0,width=180mm]{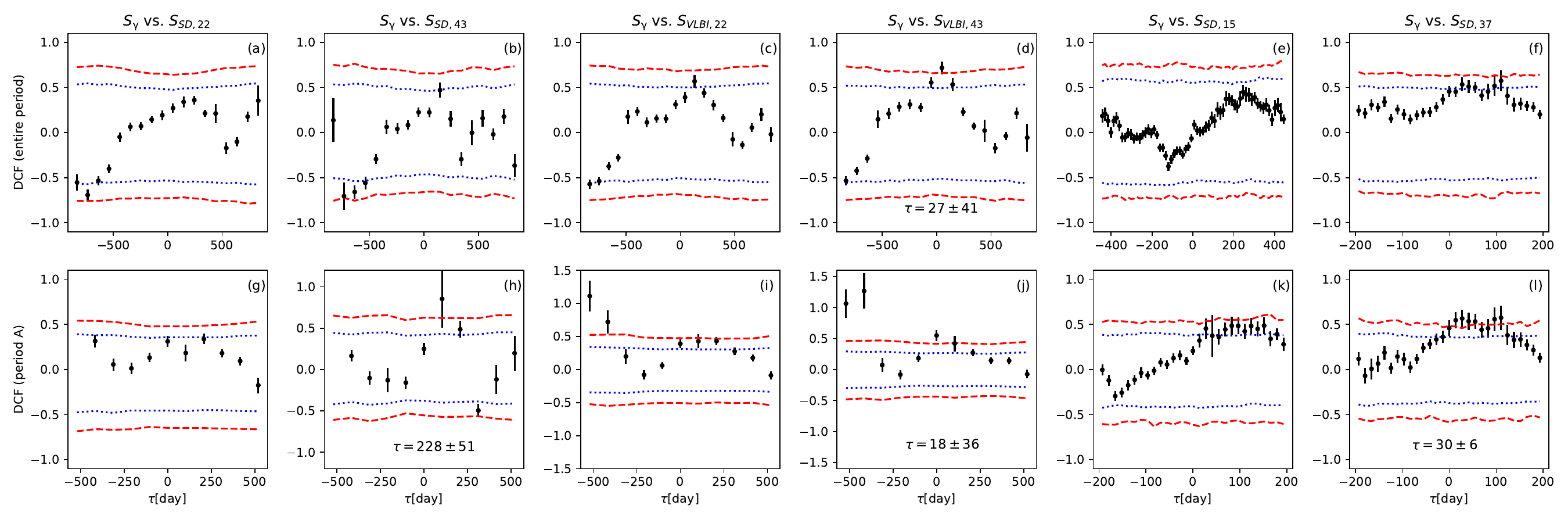}
\caption{Results of DCF analysis for gamma-ray light curve versus single-dish and VLBI core light curves over the entire period (upper panel) and period A (lower panel). The red dashed and blue dotted lines denote $3\sigma$ (99.7\%) and $2\sigma$ (95\%) confidence intervals, respectively. Positive time lags indicate that the flux density at gamma-ray leads the flux density at lower energy bands.
\label{fig:DCF results}}
\vspace{0mm} 
\end{figure*}

\subsection{Cross-correlation analysis}
\label{sec:DCF}
To investigate the potential correlation between gamma-ray flares and radio flux variations, we employed the public Python command-line tool developed by \cite{robertson2015searching} \footnote{\url{https://github.com/astronomerdamo/pydcf}} to compute the Discrete Correlation Function (DCF), as described by \cite{edelson1988discrete}. 

The DCF is a widespread method used to probe correlations in two time series data that are unevenly sampled. The process begins by collecting the set of unbinned discrete correlations:
\begin{flalign}
\label{eq:UDCF}
&\rm{UDCF}_{i j}=\frac{\left(a_i-\bar{a}\right)\left(b_j-\bar{b}\right)}{\sqrt{\left(\sigma_a^2-e_a^2\right)\left(\sigma_b^2-e_b^2\right)}},&
\end{flalign}
where $a_i$ and  $b_i$ represent two discrete time series of the light curves, $\bar{a}$ and $\bar{b}$ are the means of the time series, and $\sigma_a$ and $\sigma_b$ are the standard deviations. $e_a$ and $e_b$ denote the measurement errors associated with each light curve. Each pair of $\left(a_i-b_i\right)$ is from the measurements of the light curves that fall within the time lag bin defined by $\tau-\Delta \tau / 2 \leq \Delta t_{i j}<\tau+\Delta \tau / 2$, where $\Delta t_{i j}=t_j-t_i$ is the pairwise lag, $\tau$ is the time lag, and $\Delta\tau$ is the bin width.

Subsequently, we average the UDCF over the N pairs of the data sets that fall within $\Delta t_{i j}$:
\begin{flalign}
\label{eq:DCF}
&\mathrm{DCF}(\tau)=\frac{1}{N} \sum \mathrm{UDCF}_{i j},&
\end{flalign}
and the standard deviation of each bin is defined as
\begin{flalign}
\label{eq:std of DCF}
&\sigma_{\mathrm{DCF}}(\tau)=\frac{1}{N-1} \sqrt{\sum \left[\mathrm{UDCF}_{i j}-\mathrm{DCF}_{i j}\right]^2}.&
\end{flalign}

A positive (negative) time lag signifies that the flux density of the light curve $a_i$ leads (lags) the light curve $b_i$. In selecting the time lag range, we aimed for a length that covers about two-thirds of the shorter light curves to ensure overlapping, despite observation gaps. The bin width was chosen to be one to two times the mean cadence of the less frequently sampled light curve. A longer bin width may cause a loss of detail, while a narrower bin width could lead to gaps in the DCF curve and loss of accuracy~\citep{liodakis2018multiwavelength}.

To assess the significance of the DCF coefficient, we simulated 10,000 artificial gamma-ray light curves with similar power spectral densities (PSDs) and probability density functions (PDFs) to the real curve. The simulation was conducted using public code provided by \cite{connolly2016delightcurvesimulation} \footnote{\url{https://github.com/samconnolly/DELightcurveSimulation}}, developed based on the method described in \cite{emmanoulopoulos2013generating}. 

The program initially derives the periodogram from the observed light curve and estimates its underlying power spectral density (PSD) by fitting it to a specified PSD model. Subsequently, it constructs a histogram of the observed light curve and evaluates it using a designated probability density function (PDF) model. Based on the best-fitting PSD of the original data, a simulated light curve with N values, distributed normally, is generated. Additionally, a sequence of N pseudo-random numbers is produced from the best-fitting PDF, forming a white noise dataset. Adjustments in both spectral and amplitude are then applied to the discrete Fourier transform (DFT) of this simulated light curve. Through iterative adjustments of the amplitude and shape of the simulated curve, combined with random variations and specific models, the program  produces a simulated light curve that statistically mirrors the observed light curve. The time intervals of the light curves were increased by around 10 percent considering the aliasing effects, and to properly account for the effects of red noise leakage, the total length of simulated light curve is extended much longer than the original data and the converged final light curve is truncated to the desired length. During the simulation, a power-law of the form ${\rm PSD} \propto 1/\nu^{\beta}$ is fitted to the PSD of the light curves, where $\beta$ is a slope of the power-law spectrum. With regard to the PDFs, an uniform distribution or a log-normal distribution of the form ${\rm PDF} = (1/x\sigma \sqrt{2\pi})e^{-({\rm ln}(x)-\mu)^2/2\sigma^2}$ is fitted to the PDFs of the radio light curves, where $\mu$ and $\sigma$ are scale and shape parameters, respectively. A gamma distribution of the form ${\rm PDF} = x^{\kappa-1}({\rm \Gamma}(\kappa)\theta^{\kappa})e^{-x}$ is fitted to the gamma-ray light curve, where $x$ is flux density in a PDF, $\kappa$ is a shape parameter, $\rm \Gamma$ is a Gamma function, and $\theta$ is a scale parameter. Mean cadences were used to bin the light curves, which is about 10~days in the well-sampled OVRO, Metsähovi and Fermi LAT data, and about 46-79~days in the KVN light curves.

Cross-correlations were calculated between the artificial light curves, yielding 10,000 DCF results. The 95.45th and 99.73th percentiles of the obtained DCF coefficients' distributions defined the $2\sigma$ and $3\sigma$ confidence levels.

The time lag corresponding to the peak of the DCF result is identified as the time delay between the two light curves. To quantify the uncertainty in this lag, we employed a model-independent Monte Carlo approach, as described in \cite{peterson1998uncertainties}. We draw 10,000 random samples from a normal distribution for each data point of the light curve, using the flux density as the mean and its uncertainty as the standard deviation. From these samples, we randomly selected a value for each data point to create an artificial light curve as a subset. This process was repeated to generate a total of 10,000 subsets yielding 10,000 time lags after DCF analysis. Based on these time lags, we obtained a cross-correlation peak distribution. The mean and standard deviation of the distribution of the DCF peaks are considered to represent the time lag, $\tau$, and its uncertainty, $\sigma_{\tau}$, assuming a normal distribution~\citep{lee2017interferometric}.

We performed the cross-correlation analysis for the entire period on $S_{\rm \gamma}-S_{\rm SD,22}$, $S_{\rm \gamma}-S_{\rm SD,43}$, $S_{\rm \gamma}-S_{\rm VLBI,22}$, $S_{\rm \gamma}-S_{\rm VLBI,43}$, as well as $S_{\rm \gamma}-S_{\rm SD,15}$ and $S_{\rm \gamma}-S_{\rm SD,37}$ for comparison. Furthermore, we conducted the analysis for a specific period (i.e., period A) for the above data pairs. The results are presented in Figure~\ref{fig:DCF results}.

We observed significant correlations ($>3\sigma$) between gamma-ray flux density and 43~GHz VLBI core flux in both period A and the entire period, with positive time lags of 18$\pm$36~days and 27$\pm$41~days respectively, as shown in Figure~\ref{fig:DCF results} (j), (d). These results indicate that the peaks of the gamma-ray emissions fall within two months before to half a month after the peak of the radio flare. Similar significant correlation ($>3\sigma$) was also found between gamma-ray flux density and 37~GHz single-dish flux in period A, with a positive time lag of 30$\pm$6~days (refer to Figure~\ref{fig:DCF results} (i)). A marginal (DCF values with large uncertainties) 3$\sigma$ DCF peak was found in $S_{\rm \gamma}-S_{\rm SD,43}$ during period A (Figure~\ref{fig:DCF results} (h)), most likely resulted from the large uncertainty of the 43 GHz single-dish data. Negative time lags were also detected in $S_{\rm \gamma}-S_{\rm VLBI,22}$ and $S_{\rm \gamma}-S_{\rm VLBI,43}$ during period A. However, this is attributed to the gamma-ray flare correlating with the initial segment of the VLBI core light curves, which is part of a broader declining flux density trend, not a radio flare peak.

\begin{figure*}[!htb]
\centering 
\label{Fig:VLBI flares 43 GHz}
\includegraphics[angle=0,width=120mm]{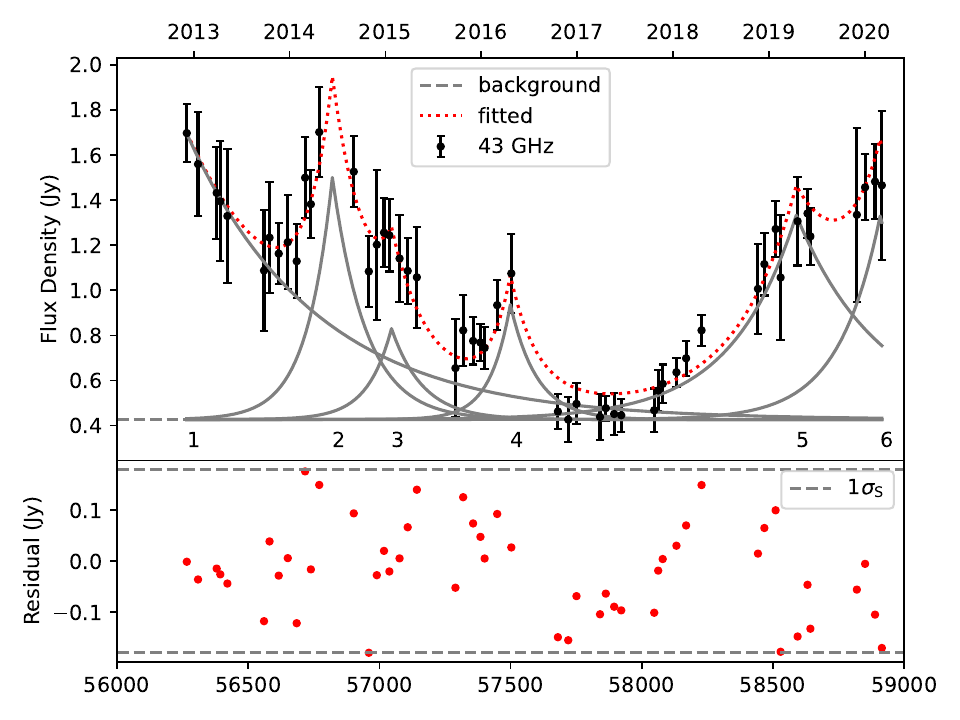}
\caption{The upper plots present the decomposed exponential flares (red dotted line) of the light curves. The gray solid lines denote the individual flares, and the gray dashed line denotes the quiescent (minimum) flux density. The red dots in the lower plots show the residuals between the data and the flare models.}
\label{Fig:radio flares}
\end{figure*}

\begin{table*}[!htb]
    \centering
    \caption{Physical parameters of the radio flares in the 43~GHz VLBI core light curve}
    \label{tab:physical params 43GHz VLBI}
    \renewcommand{\arraystretch}{1.3}
    \setlength\tabcolsep{10pt}
    \begin{tabular}{cccccccc}
        \hline
		Flare & $t_{\rm max}$ & $t_{\rm max}$ & $S_{\rm max}$ & $t_{\rm r}$ & $T_{\rm b, var}$ & $\delta_{\rm var}$ & $\theta_{\rm var}$\\ 
              & (year-month-date) & (MJD) & (Jy) & (day) & ($\times 10^{13}$ K) & & mas\\
        \hline
1 & $2012-12-06^{+42}_{-34}$ & $56267^{+42}_{-34}$ & $1.27^{+0.23}_{-0.20}$ & $380^{+23}_{-17}$ & $0.13^{+0.03}_{-0.02}$ & $5.98^{+0.40}_{-0.36}$ & $0.078^{+0.004}_{-0.004}$\\
2 & $2014-06-12^{+25}_{-17}$ & $56820^{+25}_{-17}$ & $1.07^{+0.59}_{-0.39}$ & $99^{+33}_{-42}$ & $1.68^{+1.29}_{-1.56}$ & $13.88^{+3.55}_{-4.29}$ & $0.047^{+0.008}_{-0.009}$\\
3 & $2015-01-23^{+82}_{-48}$ & $57045^{+82}_{-48}$ & $0.40^{+0.31}_{-0.30}$ & $92^{+43}_{-16}$ & $0.73^{+0.88}_{-0.60}$ & $10.52^{+4.22}_{-2.89}$ & $0.033^{+0.010}_{-0.008}$\\
4 & $2016-04-20^{+18}_{-35}$ & $57498^{+18}_{-35}$ & $0.51^{+0.28}_{-0.32}$ & $78^{+15}_{-33}$ & $1.29^{+0.95}_{-1.35}$ & $12.7^{+3.14}_{-4.45}$ & $0.034^{+0.007}_{-0.008}$\\
5 & $2019-04-15^{+18}_{-35}$ & $58588^{+18}_{-35}$ & $0.91^{+0.19}_{-0.21}$ & $246^{+18}_{-30}$ & $0.23^{+0.06}_{-0.08}$ & $7.14^{+0.67}_{-0.82}$ & $0.06^{+0.005}_{-0.005}$\\
6 & $2020-02-29^{+72}_{-26}$ & $58908^{+72}_{-26}$ & $0.90^{+0.59}_{-0.36}$ & $158^{+25}_{-55}$ & $0.55^{+0.28}_{-0.45}$ & $9.59^{+1.62}_{-2.59}$ & $0.052^{+0.007}_{-0.009}$\\
		\hline
    \end{tabular}
\tabnote{
\emph{Notes.} $t_{\rm max}$ is the time of the maximum amplitude of the flare in year-month-day and modified Julian date, $S_{\rm max}$ is the maximum amplitude of the flare in Jy, $t_{\rm r}$ is the rising time-scale in days, $T_{\rm b, var}$ is the variability brightness temperature in $10^{13}$~K, $\delta_{\rm var}$ is the variability Doppler factor, and $\theta_{\rm var}$ is the angular size of the emission region in mas.
}
\end{table*}

\subsection{Variability time scales}

\subsubsection{Radio flare decomposition}
In Section \ref{sec:DCF}, we observed strong correlations between gamma-ray flux density and 43~GHz VLBI core flux. To examine the physical properties of the specific flare that is associated to the gamma-ray flare, we performed radio flare decomposition. 

The variations in radio flux density observed in compact sources are typically characterized by a combination of multiple flare components. These components can be effectively modeled using exponential functions~\citep{valtaoja1999total}:
\begin{flalign}
\label{eq:flare model}
&\Delta S(t) = 
\begin{cases}
\Delta S_{\rm max} e^{(t-t_{\rm max}) / {t_{\rm r}}}, &\space t<t_{\rm max},\\ 
\Delta S_{\rm max} e^{(t_{\rm max}-t) / 1.3t_{\rm r}}, &\space t>t_{\rm max},
\end{cases}
\end{flalign}
where $S_{\rm max}$ is the maximum amplitude of a flare in Jy, $t_{\rm max}$ is the time of when the flare reaches its maximum in MJD, and $t_{\rm r}$ is the rise time-scale of the flare in days. 

Prior to the decomposition, a running average was performed to identify the radio flaring periods. We use a window size of 5 data points, slide this window across the light curve data set from the beginning to the end. For each window position, calculate the average of the data points within that window. Move the window one data point over and repeat, creating a new series of averages that represent the smoothed data. The radio flaring periods are then defined as periods when the flux density was above average. We defined the lowest observed flux level of the light curve (i.e., 0.43~Jy at 43~GHz for VLBI core light curves) as the constant quiescent flux level and subtracted it before fitting the flares. Subsequently, we fitted the flare model defined in Equation~(\ref{eq:flare model}) to the first prominent peak identified from the running average result, and subtracted the fitted flare flux. We repeated the procedure until the residual was within the range of $3\sigma_{\rm S}$ for the single-dish data and $1\sigma_{\rm S}$ for the VLBI core flux, where $\sigma_{\rm S}$ represents the root mean square of the light curve's statistical flux uncertainty~\citep{kim2022magnetic}. The initially obtained parameters are subsequently constrained using the Markov Chain Monte Carlo (MCMC) method, implemented through a Python package~\citep{foreman2013emcee} \footnote{\url{https://emcee.readthedocs.io/en/stable/}}. The uncertainties of the parameters are from the $1\sigma$ confidence intervals of the distributions of the parameters.The results of the decomposition are shown in Figure~\ref{Fig:radio flares}. The best-fit parameters are summarized in Table~\ref{tab:physical params 43GHz VLBI}.

\subsubsection{Physical parameters}
Utilizing the variability time scales obtained in the previous section, we can estimate the variability brightness temperature, $T_{\rm b}^{\rm var}$, the variability Doppler factors, $\delta_{\rm var}$, and the angular size, $\theta_{\rm var}$, of the emission region. These estimations are based on the assumption that the flux density variability is intrinsic to the source, and the variable component is spherical with a Gaussian brightness distribution~\citep{fuhrmann2008testing, kang2021interferometric}. The following functions are used for these calculations:
\begin{flalign}
\label{eq:physical parameters}
&T^{\rm var}_{\rm b} = 4.09 \times 10^{13} \left(\frac{D_{\rm L}}{\nu~t_r}\right)^2 \frac{\Delta S}{(1+z)^4} \\
&\delta_{\rm var} = (1+z)\left(\frac{T^{\rm var}_{\rm b}}{T_{\rm b,eq} }\right)^{1/3} \\
&\theta_{\rm var} = (1+z)\frac{c~t_{\rm r}}{D_{\rm L}} \delta_{\rm var}, &     
\end{flalign}
where $\Delta S$ is the flux difference between the peak and the peak/$e$ of a flare measured in Jy, $\nu$ is the observing frequency in GHz, $z=0.996$ is the redshift~\citep{albareti201713th}, $D_{\rm L}$ is the luminosity distance with $D_{\rm L}=6755.2$~Mpc, assuming ${\rm H}_0=67.8$~km/sec/Mpc, ${\rm \Omega_{matter}}=0.308$, and ${\rm \Omega_{vaccuum}}=0.692$~\citep{wright2006cosmology}, $t_{\rm r}$ is the rise time-scale of the flare in days. In the calculation of the Doppler factor, we assume that the equipartition brightness temperature $T_{\rm b,eq}= 5\times 10^{10}$~\citep{readhead1994equipartition}. The estimated physical parameters of the flares in 22 and 43~GHz single-dish and VLBI light curves are listed in Table~\ref{tab:physical params 43GHz VLBI}. The uncertainties were determined using a Python package developed by Eric O. Lebigot, specifically designed for calculations involving uncertainties based on error propagation theory\footnote{\url{http://pythonhosted.org/uncertainties/}}.

\section{Discussion}

\subsection{Variability of the Spectral Index}
In our study of  blazar 1308+326, we noticed a correlation between the time variation of the single-dish spectral indices and the gamma-ray light curve. This correlation is characterized by an increase in spectral indices before the notable gamma-ray flare on April 26, 2014, followed by a decrease after the flare. This phenomenon could be driven by multiple physical mechanisms, including synchrotron self-absorption (SSA) and the impact of shock waves in the jet.

In the radio regime (e.g., 15-43~GHz), the primary radiation mechanism for AGN jets is synchrotron radiation, produced by charged particles moving at relativistic speeds in a magnetic field. These particles usually follow a power-law energy distribution, resulting in higher flux and brightness temperature at lower frequencies. However, when the brightness temperature exceeds a threshold, disrupting the thermal equilibrium, the system increases its absorption rate to restore balance. This in turn makes the synchrotron source optically thick below a certain critical frequency~\citep{condon2016essential}.

The observed increase in spectral indices ($\alpha_{22-43}\sim0,\alpha_{15-22}>0$) suggest the emission region at 22-43~GHz has become optically thick due to some perturbation (e.g., a shock) in the synchrotron emission region, possibly causing the absorption of low-energy photons and yielding an inverted spectrum, as shown in the synchrotron self-absorbed region (e.g., the 43~GHz radiation is less absorbed than the 22~GHz radiation).

The shock-in-jet model is often used to explain these observations. According to \cite{blandford1979relativistic}, the instabilities and non-steady motion within the jet form a shock wave near the starting point of the jet. The moving shock wave compresses the local magnetic field and particles, thereby increasing the local magnetic field strength and particle density within a thin layer behind the shock front. These relativistic particles may engage in IC with surrounding photons, up-scattering lower-energy photons to gamma-ray energy levels, and thus triggering a gamma-ray flare.

During the propagation of the shock downstream the jet (where the jet is optically thin at radio frequency, e.g., 22-43~GHz), the synchrotron particle density increases and the particle energy distribution shifts to higher energy (i.e., particle acceleration), causing strong absorption of the synchrotron radiation at the radio regime and turning the spectral index positive ($\alpha_{22-43}\sim0,\alpha_{15-22}>0$). As described earlier, the shocked particles lose their energy via IC. After losing energy, these particles continue to generate synchrotron radiation, causing the spectral index to decline back into negative values. Ultimately, these particles cool further due to adiabatic expansion, leading to a continued decrease in the spectral index.

\begin{figure*}[!htb]
\centering
\includegraphics[angle=0,width=180mm]{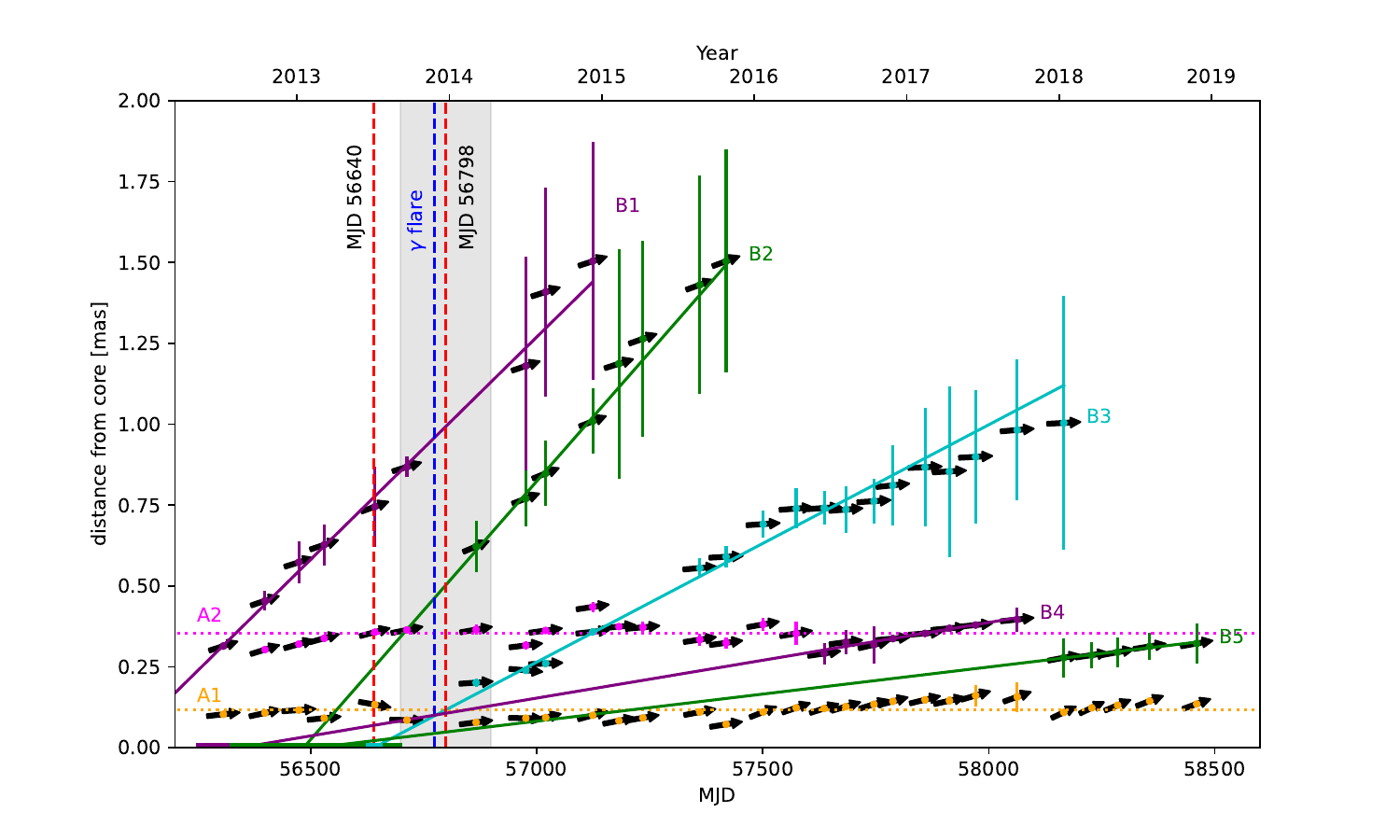}
\caption{Separation vs. time of the components in the jet of the 1308+326 from the core. Purple, green, and cyan dots denote bright, moving knots, whereas orange and pink dots are associated with stationary components. Error bars on each measurement reflect the approximate $1\sigma$ positional uncertainties, derived from observed brightness temperatures. Horizontal dotted lines represent the average positions of stationary components A1 and A2. The core component is expected to remain at the x-axis. Solid oblique lines represent the linear fits to the components’ motions, and extrapolate the components' motion back to the epochs of ejection times. The horizontal lines along x-axis indicates the uncertainties of the ejection times of the jet components. Red vertical dashed lines, ordered from left to right, mark the possible moments when component B3 emerges from the core and intersects with stationary components A1 and A2. The gray shaded region indicates the approximate duration required for B3 to pass through A1.
 \label{fig:jet components}}
\vspace{0mm} 
\end{figure*}

\begin{figure*}[!htb]
\centering
\includegraphics[angle=0,width=180mm]{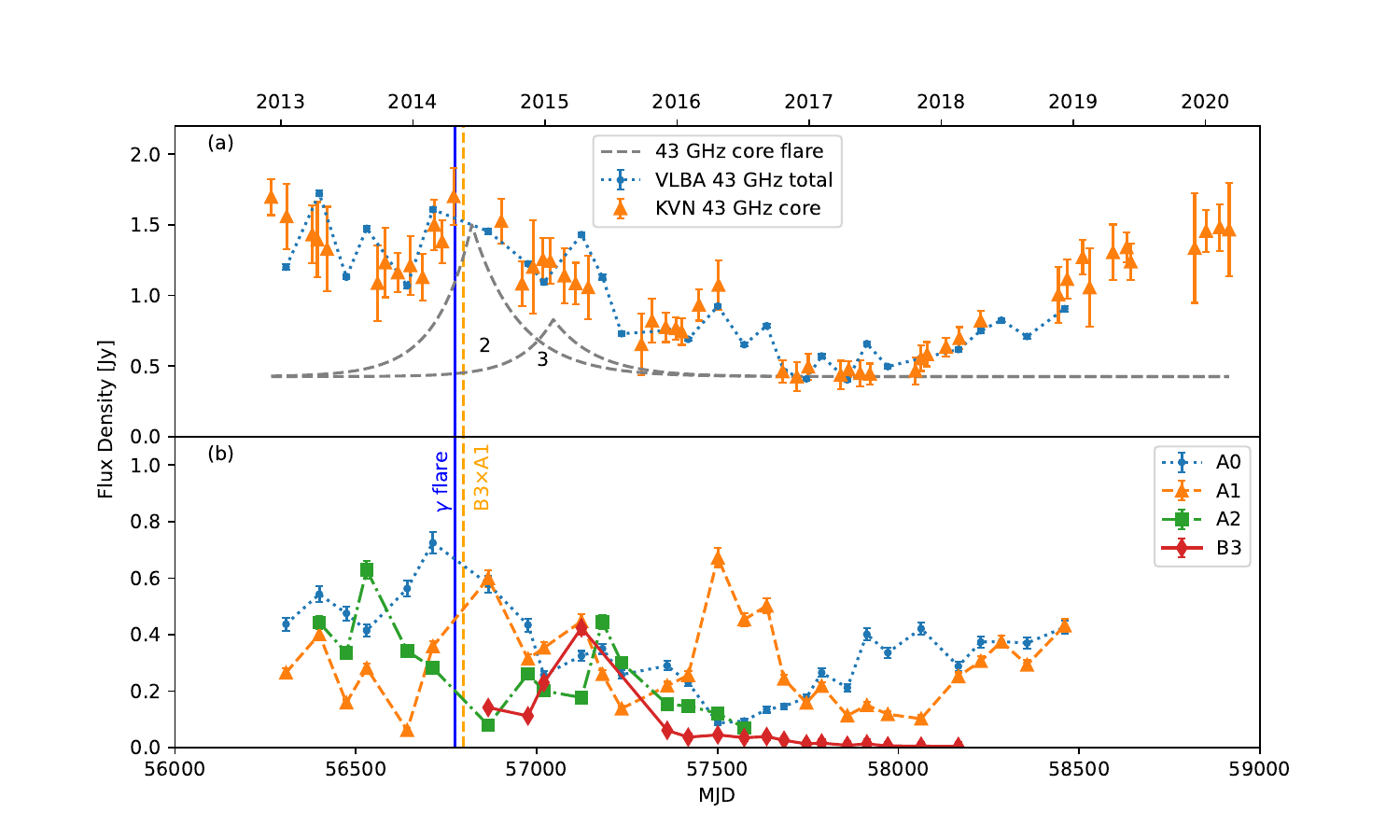}

\caption{The upper plot presents the total flux density of the 43~GHz VLBA observation, represented by blue dots, as well as the core flux density at 43~GHz from KVN, represented by an orange triangle. The gray dashed lines represent Flare 2 and Flare 3, which were identified through radio flare decomposition. The lower plot presents the flux density of each component identified in VLBA data. A vertical solid line marks the peak time of the gamma-ray flare that occurred in 2014. Vertical dashed lines mark the potential times when component B3 passes stationary components A1 and A2.
 \label{fig:BU flux}}
\vspace{0mm} 
\end{figure*}

\subsection{Connection between Gamma-ray emission and Radio Flares}
\label{sec:connection}
In section \ref{sec:DCF}, we observed a $>3\sigma$ correlation between gamma-ray flux and the 43~GHz VLBI core flux, with a positive time lag of $27\pm41$~days. This strong correlation indicates the gamma-ray emission may occur within 68 days before to 14 days after the peak of the radio core flares. Through the decomposition radio core flares, we found that the time lag between the peak time of the Flare 2 in the 43~GHz VLBI core light curve (MJD~$56820^{+25}_{-17}$) and the 2014 gamma-ray flare (MJD~56773) falls within the error margin, aligning with the time lag deduced from the DCF analysis.

\cite{leon2011connection} emphasized that correlation analyses often focus on the distance between peaks, especially when flares have different timescales. In light of this, we consider it crucial to compare the onset of radio and gamma-ray flares rather than their peaks. We define the beginning of a radio flare as the epoch when its flux reaches 1/e of the maximum flux, namely:
\begin{flalign}
\label{eq:radio flare onset time}
&t_0 = t_{\rm max} - t_{\rm r}, &
\end{flalign}
where $t_{\rm max}$ is the peak time of the radio flare and $t_{\rm r}$ is the variability timescale~\citep{lahteenmaki_testing_2003}, also representing the rise time needed for a flare to develop from $S_{\rm max}/e$ to $S_{\rm max}$. Substituting the parameters ($t_{\rm max}=56820^{+25}_{-17}$ in MJD and $t_{\rm r}=99^{+33}_{-42}$~days) into the equation~(\ref{eq:radio flare onset time}) we found that Flare 2 in the 43~GHz core light curve started rising on MJD~$56721^{+41}_{-45}$, about 52 days before the 2014 gamma-ray flare. This aligns with the findings of \cite{leon2011connection} regarding Fermi/LAT blazars, which state that strong gamma-ray flares tend to occur shortly after the onset of mm radio flares. This suggests that the gamma-ray emission in blazars originates from the same disturbances in the relativistic jet (i.e., shocks) that produce the radio flares. In the case of 1308+326, we can infer that the emission region of the 2014 gamma-ray flare is located downstream of the 43~GHz radio core.

On the other hand, studies have indicated that the emergence of new VLBI components is often associated with radio flares~\citep{jorstad_multiepoch_2001}. Therefore, we examined the evolution of jet components during the same period using high resolution VLBA data. Indeed, the on set time of Flare 2 coincides with the ejection time of a new jet component.

Figure~\ref{fig:jet components} presents the motion of the jet components versus time at 43~GHz (reproduced using published VLBA data from \citealp{weaver2022kinematics})]. We observed that the KVN VLBI 43~GHz core Flare 2 started rising on MJD~$56721^{+41}_{-45}$, 81 days alfter the ejection of the VLBA jet component B3 on MJD~$56640^{+24}_{-25}$. More over, the interception time of component B3 through the stationary feature A1 (MJD~$56798$) coincided with the the peak time of 2014 gamma-ray flare (MJD~$56773$), followed by the peak time of Flare 2 (MJD~$56820^{+25}_{-17}$). 

Combined with the previous spectral analysis, one possible scenario that may explain the underlying mechanism of the 2014 gamma-ray flare's origin is as follows:
When the moving shock wave propagates through the relativistic jet and reaches the radio core region, which is characterized as a conical standing shock (Marscher, 2008), a new jet component, B3, emerges. Directly in the path of B3 is a stationary feature, A1, which may represent another standing shock resulting from the recollimation of the jet~\citep{fromm2015recollimation}. As B3 passes through A1, the radio core emission begins to rise. The accelerated electrons up-scatter the surrounding photons via the inverse Compton (IC) mechanism, causing a gamma-ray burst. As B3 moves further away, the radio flare reaches its peak and then begins to decline. Referencing the angular size of B3 and A1, ($a_{\rm B3} = 0.086\pm0.019~{\rm mas}, a_{\rm A1} = 0.079\pm0.014~{\rm mas}$), as well as the proper motion of B3, ($\mu_{\rm B3}=0.278 \pm 0.012~{\rm mas~yr^{-1}}$), we can roughly calculate that the time required for B3 to pass through A1 is approximately 108 days around the epoch of their trajectory intersection (MJD 56798). In Figure~\ref{fig:jet components}, we have highlighted this range in gray.

It is possible that the trajectory of B3 in three-dimensional space does not overlap with A1. However, an examination of the total flux density and individual component flux variations in the 43 GHz VLBA data reveals a close relationship between them.

In the upper panel of Figure~\ref{fig:BU flux}, a consistency in the overall trends of the VLBA 43~GHz total flux density with that of the KVN 43~GHz core flux can be observed. Notably, a pronounced increase in the VLBA total flux density was identified shortly before the gamma-ray flare. This increment was predominantly driven by flux density variations in component A0, which represents the core in the VLBA image (lower panel of Figure~\ref{fig:BU flux}).
A temporal correlation was established between the timing of the gamma-ray flare and the calculated moment when the moving component B3 crossed the stationary component A1, coinciding with an increase in the flux of A1. The surge in radio emission could be attributed to the reactions between the materials in B3 and A1.
Gamma-ray emissions, which are optically thin, were detected before the new component became visible upon crossing optically thick regions. It can be observed from lower panel of Figure~\ref{fig:BU flux} that after the occurrence of the gamma-ray flare, the flux of A1 reached its peak, coinciding with the time when B3 became visible.

\subsection{Location of Gamma-ray Emission Region}
\label{sec:loc}
In the previous sections, we have identified a strong correlation between 2014 gamma-ray flare and the KVN 43~GHz core Flare 2. Furthermore, It is highly possible that the radio flare was induced by a moving shock, the jet component B3, which could also be responsible for the gamma-ray flare. Assuming that the speed of component B3 represents the speed of the moving shock and remains constant while propagating, we can estimate the location of the gamma-ray emission based on the timing of the gamma-ray flare’s peak, the ejection time of component B3, and the position of the radio core.

First, calculate the observed time delay between the peak time of the gamma-ray flare, $t_{\rm \gamma}$, and the ejection time of component B3, $t_{\rm B3}$:
\begin{flalign}
\label{eq:delta t}
&\Delta t_{\rm obs}=t_{\rm \gamma}-t_{\rm B3},&
\end{flalign}
where $t_{\rm B3} = 56640^{+24}_{-25}$ is the extrapolated epoch of zero separation between the component B3 and the core.

Then, convert this time difference into the distance between the radio core and the gamma-ray emission region in the source frame:
\begin{flalign}
\label{eq:distance between gamma-ray region and radio core}
&\Delta r[m] = \frac{\beta_{\rm app}c \Delta t_{\rm obs}}{(1+z){\rm sin} \theta},&
\end{flalign}
where $\beta_{\rm app}c$ is the apparent jet speed of B3, and $\theta$ is the jet viewing angle of B3 in degree~\citep{kramarenko_decade_2021}.

Next, estimate the location of the radio core relative to the central engine. As the radio core locates much further than the jet apex from the central engine, the location of the radio core in the jet, $r_{\rm c}$, is approximately equal to the distance of the radio core to the central engine, and can be estimated as follows:
\begin{flalign}
\label{eq:radio core location}
&r_{\rm c} [\rm pc]= \frac{\Omega_{r\nu}}{\nu^{1/k_{\rm r}}sin \theta},&
\end{flalign}
where $\nu$ is the observed frequency in GHz, $k_{\rm r}$ is a power-law index that characterizes the frequency dependence of the position of the radio core~\cite{lobanov1998ultracompact}. The value of $k_{\rm r}$ depends on the electron energy spectrum, the magnetic field and the particle density distributions. Studies on the core shift effect in the parsec-scale jet of other AGNs reveal that $k_{\rm r} \approx 1$~\citep{hada2011origin, mohan2015frequency, paraschos2023multi}, indicating an equipartition between the energy densities of jet particles and the magnetic field. $\Omega_{r\nu}$ is the core shift measure defined in \cite{lobanov1998ultracompact} as follows:
\begin{flalign}
\label{eq:core shift measure}
&\Omega_{r\nu}=4.85\cdot 10^{-9}\frac{\Delta r_{\rm mas} D_{\rm L}}{(1+z)^2}\frac{\nu_1^{1/k_{\rm r}} \nu_2^{1/k_{\rm r}}}{\nu_2^{1/k_{\rm r}}-\nu_1^{1/k_{\rm r}}},&
\end{flalign}
where $\Delta r_{\rm mas}$ is the core shift in milliarcseconds when measured at two frequencies, $\nu_1$ and $\nu_2$ ($\nu_1 < \nu_2$). Ideally, $\Omega_{r\nu}$ is constant for all frequency pairs.

We used the median core shift value of 0.78~mas from \cite{plavin2019significant}, which was  measured between 2 and 8~GHz. Then $\Omega_{r2{,}8}$ was calculated as 17.10~pc~GHz. We adopted the physical parameters of jet component B3 from \cite{weaver2022kinematics}, yielding $\theta=1.34\pm0.24$~deg and $\beta_{\rm app}=14.52\pm0.65$. 

Finally, by substituting all these parameters into the equations~(\ref{eq:delta t})--(\ref{eq:core shift measure}), we calculated the 43~GHz radio core position as $r_{\rm c,43GHz}=17.0\pm3.0$~pc, and the distance from the core to the gamma-ray emission region $\Delta r$ to be $34.89^{+8.99}_{-9.17}$~pc. These results position the gamma-ray flare emission region $r_{\rm \gamma}$ at $51.89^{+11.32}_{-11.47}$~pc ($16.01^{+3.49}_{-3.54}\times10^{19}$ cm). Hence, the gamma-ray flare occurred 40--63 parsecs away from the central engine.

\begin{figure*}[!htb]
\centering
\includegraphics[angle=0,width=180mm]{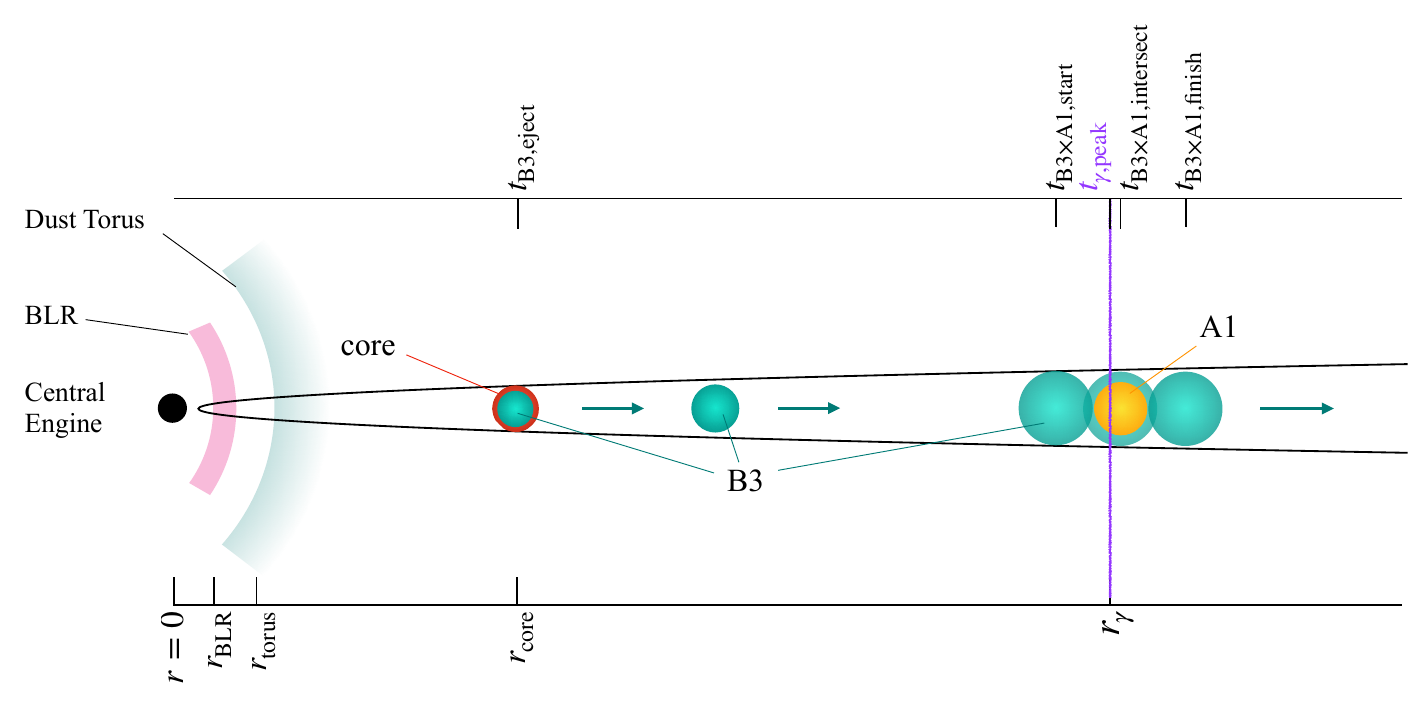}

\caption{A schematic view for the origin of the 2014 gamma-ray flare. $t_{\rm B3,eject} = {\rm MJD}~56640^{+24}_{-25}$ corresponds to the ejection time of the newly emerged jet component B3 from the radio core. As B3 propagates downstream, it encounters and passes through a stationary jet component, A1. The time sequence of this process is denoted by $t_{\rm B3\times A1, start}={\rm MJD}~56690$, $t_{\rm B3\times A1, intersect}={\rm MJD}~56798$ and $t_{\rm B3\times A1, finish}={\rm MJD}~56906$.
During this process, a gamma-ray flare bursts out, with the peak time of this flare denoted as $t_{\rm \gamma, peak} = {\rm MJD}~56773$. The axis at the bottom of the diagram shows the distance from the central engine. $r_{\rm BLR}\sim 1~{\rm pc}$ and $r_{\rm torus}\sim 2~{\rm pc}$ denote the approximate distances of the BLR and dust torus, respectively. $r_{\rm core}=17\pm3~{\rm pc}$ is the position of the 43~GHz radio core, and $r_{\rm \gamma}=52\pm11~{\rm pc}$ is the position of the gamma-ray emission region we calculated based on the velocity of B3.
 \label{fig:diagram}}
\vspace{0mm} 
\end{figure*}

\subsection{Seed photon candidates}
It is generally believed that the gamma-ray emission from AGNs is caused by IC scattering, which requires (1) a large number of high-velocity particles and (2) seed photons. In \ref{sec:connection}, we posited that a large number of high-velocity particles may come from electrons accelerated by the moving hocks. But where do the seed photons come from?

Among the candidates for seed photons, non-thermal photons may come from the synchrotron jet itself, causing SSC scattering. Meanwhile, thermal photons may originate from the BLR, the dust torus, or CMB radiation. By comparing the location of gamma-ray emission with the distances to these photon sources, we can further narrow down the origin of the seed photons.

Assuming the BLR is a spherically symmetric shell surrounding the SMBH, the radius of the BLR, $r_{\rm BLR}$, is roughly proportional to the square root of the accretion disk luminosity, $L_{\rm disc}$~\citep{wandel1999central}, i.e., $r_{\rm BLR}[{\rm cm}] = 10^{17} (L_{\rm disc} / 10^{45}~{\rm erg~s^{-1}})^{0.5}$, where $L_{\rm disc}=9\times 10^{45}~{\rm erg~s^{-1}}$ for 1308+326~\citep{ghisellini2015fermi}.

The dust torus, on the other hand, might be located at a distance $r_{\rm torus}$ from the central black hole, and $r_{\rm torus}[{\rm cm}]= 2\times 10^{18} (L_{\rm disc} / 10^{45}~{\rm erg~s^{-1}})^{0.5}$~\citep{ghisellini2015fermi}.

Calculations in section \ref{sec:loc} show that the emission region of the 2014 gamma-ray flare in 1308+326 is approximately at a distance $16\times10^{19}~{\rm cm}$ from the central engine, which is far outside the BLR ($3\times 10^{17}~{\rm cm}$), and even away from the dust torus ($6\times 10^{18}~{\rm cm}$). Therefore, the seed photons of the IC for the gamma-ray flare are less likely to originate from the BLR, and more likely originate either from the jet itself or from the dusty torus, or from CMB radiation.

It should be noted that the sizes of the BLR and dust torus mentioned above are rough estimates based on specific emission models. These estimates are subject to constraints imposed by varying accretion rates, jet orientations, and environmental conditions, indicating that the actual scenario is likely to be more complex. The conclusions drawn should be approached with a degree of caution due to the absence of more specific data.  We look forward to subsequent research that can provide more detailed observations and data to help validate or refine our current assumptions.

\subsection{Brightness Temperature and Emission Region}
Based on our analysis, we have estimated the variability brightness temperature and the size of the emission region. The extremely high brightness temperature suggests that we are likely observing a relativistic emission region. This indicates that the radiation we are capturing is generated by particles moving at relativistic speeds. Furthermore, the size of the estimated 43~GHz variable emission region ranges from 0.033 to 0.078~mas. This small size indicates that the observed variations originate from a highly compact area, which is consistent with the relativistic nature suggested by the high brightness temperature.

\section{Summary}
In this study, we have conducted a comprehensive multi-frequency analysis of the blazar 1308+326, covering the period from December 2012 to March 2020. We utilized single-dish and VLBI radio observations at 22 and 43~GHz, along with gamma-ray data from the Fermi LAT, and single-dish data from OVRO and Metsähovi. Spectral analysis of the single-dish light curves indicated that the source underwent an optically thin to thick transition before the 2014 gamma-ray flare and a thick to thin transition after the flare, suggesting a change in the jet’s physical characteristics. This spectral evolution is consistent with the shock-in-jet model, where a moving shock increases the density of particles.

Our analysis also identified significant correlations between gamma-ray emissions and the 43~GHz VLBI core emission, with a time lag of approximately 27~days. This suggests that gamma-ray flares, particularly the notable early 2014 flare, occurred roughly 50~days after the onset of a VLBI core flare, with the emission region located 40-63~parsecs from the central engine. Considering estimates of the sizes of the BLR and the dust torus, it seems more plausible that the seed photons for IC scattering, which triggered the gamma-ray flare, originated from the jet itself, the dusty torus, or the CMB radiation, rather than the BLR.

The emission from the 2014 gamma-ray flare likely occurred as a shock moved through a stationary feature in the jet downstream from the radio core. The moving shock emerged from the core as a newly ejected component, compressed the material in its trajectory, enhanced the electrons' energy and led to the inverse IC scattering of low-energy photons, resulting in a burst of gamma-ray flare. The temporal and spatial relationships of these events are illustrated in Figure~\ref{fig:diagram}

As the shock moves through the radio core, it also orders the magnetic field, leading to observable changes in polarization. We hope that future research will further elucidate these dynamics and their implications for our understanding of blazar emissions.


\acknowledgments

We are grateful to the staff of the KVN who helped to operate the array and to correlate the data. The KVN is a facility operated by KASI (Korea Astronomy and Space Science Institute). The KVN observations and correlations are supported through the high-speed network connections among the KVN sites provided by KREONET (Korea Research Environment Open NETwork), which is managed and operated by KISTI (Korea Institute of Science and Technology Information). This work has made use of Fermi-LAT data supplied by \citet{abdollahi2023}, \url{https://fermi.gsfc.nas a.gov/ssc/data/access/lat/LightCurveRepository}. 
This work was supported by a National Research Foundation of Korea (NRF) grant funded by the Korean
government (MIST) (2020R1A2C2009003).

\bibliography{paper}





\newpage

\appendix
\section{}
\label{Append_A}
\renewcommand{\thetable}{\Alph{section}\arabic{table}}
\setcounter{table}{0}
\begin{table*}[!htb]
    \small 
    \centering
    \caption{Comparison of the Four Measurements with Large Errors and Other Data}
    \label{tab:comparison}
    \renewcommand{\arraystretch}{1}
    \setlength\tabcolsep{6pt}
    \begin{tabular}{ccccccc | ccccccc}
\hline
\hline
    &   &   & 22~GHz &   &   &   &   &   &   & 43~GHz &   &   &   \\ 	
\hline
  MJD$_{\rm avg}$ & S$_{\rm avg}$ & S$_{\rm err}$ & MJD & TELE & res$_{\rm AZ}$ & res$_{\rm EL}$ &MJD$_{\rm avg}$ & S$_{\rm avg}$ & S$_{\rm err}$ & MJD & TELE & res$_{\rm AZ}$ & res$_{\rm EL}$\\ 
          & (Jy) & (Jy) &   &  & (Jy) & (Jy) &   & (Jy) & (Jy) &  &  & (Jy) & (Jy)\\
\hline
56959 &     1.89 &     0.05 &     56959.0897 &     KYS &     0.24 &     0.25 &     {\color{red}56960} &     {\color{red}2.48} &     {\color{red}0.25} &     {\color{red}56960.0556} &     {\color{red}KTN} &     {\color{red}0.68} &     {\color{red}0.47} \\
 &      &      &     56959.1303 &     KUS &     0.2 &     0.17 &      &      &      &      &      &      &      \\
 &      &      &     56959.1702 &     KYS &     0.25 &     0.26 &      &      &      &      &      &      &      \\
 &      &      &     56959.1704 &     KTN &     0.22 &     0.27 &      &      &      &      &      &      &      \\
 &      &      &     56959.1704 &     KUS &     0.19 &     0.25 &      &      &      &      &      &      &      \\
 &      &      &     56960.0207 &     KUS &     0.39 &     0.31 &      &      &      &      &      &      &      \\
 &      &      &     56960.0555 &     KUS &     0.2 &     0.26 &      &      &      &      &      &      &      \\
\hline
56990 &     2.13 &     0.05 &     56989.9352 &     KYS &     0.42 &     0.67 &     {\color{red}56990} &     {\color{red}2.56} &     {\color{red}0.24} &     {\color{red}56989.9747} &     {\color{red}KUS} &     {\color{red}0.46} &     {\color{red}0.46} \\
 &      &      &     56989.9746 &     KTN &     0.19 &     0.2 &      &      &      &      &      &      &      \\
 &      &      &     56989.9747 &     KUS &     0.32 &     0.36 &      &      &      &      &      &      &      \\
\hline
57017 &     2.19 &     0.01 &     57017.922 &     KYS &     0.17 &     0.16 &     57017 &     1.83 &     0.06 &     57016.9648 &     KUS &     0.14 &     0.21 \\
 &      &      &     57017.8887 &     KUS &     0.15 &     0.15 &      &      &      &     57016.9648 &     KTN &     0.5 &     0.38 \\
 &      &      &     57017.8887 &     KYS &     0.14 &     0.13 &      &      &      &     57016.9993 &     KUS &     0.2 &     0.14 \\
 &      &      &     57017.8591 &     KUS &     0.14 &     0.15 &      &      &      &     57016.9993 &     KTN &     0.36 &     0.29 \\
 &      &      &     57017.859 &     KTN &     0.21 &     0.16 &      &      &      &     57017.859 &     KYS &     0.29 &     0.31 \\
 &      &      &     57017.859 &     KYS &     0.17 &     0.18 &      &      &      &      &      &      &      \\
 &      &      &     57017.8887 &     KTN &     0.15 &     0.16 &      &      &      &      &      &      &      \\
 &      &      &     57016.9993 &     KTN &     0.14 &     0.14 &      &      &      &      &      &      &      \\
 &      &      &     57016.9993 &     KUS &     0.12 &     0.15 &      &      &      &      &      &      &      \\
 &      &      &     57016.9649 &     KYS &     0.14 &     0.19 &      &      &      &      &      &      &      \\
 &      &      &     57016.9648 &     KTN &     0.14 &     0.17 &      &      &      &      &      &      &      \\
 &      &      &     57016.9648 &     KUS &     0.13 &     0.15 &      &      &      &      &      &      &      \\
 &      &      &     57016.9993 &     KYS &     0.15 &     0.19 &      &      &      &      &      &      &      \\
\hline
57038 &     1.93 &     0.02 &     57037.9396 &     KYS &     0.25 &     0.25 &     57038 &     1.32 &     0.04 &     57037.8033 &     KUS &     0.17 &     0.19 \\
 &      &      &     57037.9396 &     KTN &     0.16 &     0.15 &      &      &      &     57037.8375 &     KUS &     0.16 &     0.18 \\
 &      &      &     57037.9396 &     KUS &     0.18 &     0.16 &      &      &      &     57037.9064 &     KTN &     0.32 &     0.45 \\
 &      &      &     57037.9063 &     KUS &     0.19 &     0.19 &      &      &      &      &      &      &      \\
 &      &      &     57037.9064 &     KTN &     0.17 &     0.17 &      &      &      &      &      &      &      \\
 &      &      &     57037.8375 &     KUS &     0.19 &     0.14 &      &      &      &      &      &      &      \\
 &      &      &     57037.8033 &     KTN &     0.15 &     0.17 &      &      &      &      &      &      &      \\
 &      &      &     57037.8033 &     KUS &     0.18 &     0.17 &      &      &      &      &      &      &      \\
 &      &      &     57037.8375 &     KTN &     0.19 &     0.16 &      &      &      &      &      &      &      \\
\hline
57357 &     1.26 &     0.04 &     57356.9393 &     KTN &     0.25 &     0.23 &     {\color{red}57357} &     {\color{red}2.0} &     {\color{red}0.5} &     {\color{red}57356.8984} &     {\color{red}KTN} &     {\color{red}0.64} &     {\color{red}0.47} \\
 &      &      &     57357.0199 &     KTN &     0.2 &     0.28 &      &      &      &      &      &      &      \\
 &      &      &     57357.0994 &     KTN &     0.19 &     0.22 &      &      &      &      &      &      &      \\
\hline
57385 &     1.2 &     0.03 &     57385.0335 &     KTN &     0.16 &     0.14 &     {\color{red}57385} &     {\color{red}1.67} &     {\color{red}0.11} &     {\color{red}57385.0335} &     {\color{red}KTN} &     {\color{red}0.32} &     {\color{red}0.38} \\
 &      &      &     57384.8747 &     KTN &     0.14 &     0.12 &      &      &      &      &      &      &      \\
 &      &      &     57384.9144 &     KTN &     0.18 &     0.15 &      &      &      &      &      &      &      \\
 &      &      &     57384.9937 &     KTN &     0.15 &     0.15 &      &      &      &      &      &      &      \\
\hline
57503 &     1.48 &     0.29 &     57502.5222 &     KTN &     0.24 &     0.2 &     57503 &     1.51 &     0.05 &     57502.5222 &     KTN &     0.51 &     0.41 \\
 &      &      &      &      &      &      &      &      &      &     57502.6447 &     KTN &     0.37 &     0.42 \\

		\hline
    \end{tabular}
\tabnote{
\emph{Notes.} MJD$_{\rm avg}$: average observing time in modified Julian date; S$_{\rm avg}$: weighted average flux density in Jy; S$_{\rm err}$ uncertainty of the flux density in Jy; MJD: observing time in modified Julian date; TELE: name of the antenna for the observation; res$_{\rm AZ}$: root mean square of the residual obtained after fitting a 1-D Gaussian model in the azimuth direction; res$_{\rm EL}$: root mean square of the residual obtained after fitting a 1-D Gaussian model in the elevation direction.

The four measurements with significant errors are highlighted in red.

}
\end{table*}

\end{document}